\documentclass[aps,prd,twocolumn,superscriptaddress]{revtex4-2}
\usepackage{graphicx}
\usepackage{hyperref}
\usepackage{url}

\begin{document}
\title{Multi-TeV $\gamma$-ray candidates from GRB 221009A: \\ a downturn in the intrinsic $\gamma$-ray spectrum, an echo of the prompt emission phase, and intergalactic electromagnetic cascades}
\author{Timur A. Dzhatdoev}
\email[]{timur1606@gmail.com}
\affiliation{Institute for Nuclear Research of the Russian Academy of Sciences, 60th October Anniversary Prospect 7a, Moscow 117312, Russia}
\author{Jagdish C. Joshi}
\affiliation{Aryabhatta Research Institute of Observational Sciences (ARIES), Manora Peak, Nainital 263001, India}
\affiliation{Centre for Astro-Particle Physics (CAPP) and Department of Physics, University of Johannesburg, PO Box 524, Auckland Park 2006, South Africa}
\author{Abhijit Roy}
\affiliation{Gran Sasso Science Institute, via F. Crispi 7 -- 67100, L’Aquila, Italy}
\affiliation{INFN/Laboratori Nazionali del Gran Sasso, via G. Acitelli 22, Assergi (AQ), Italy}
\author{Grigory I. Rubtsov}
\affiliation{Institute for Nuclear Research of the Russian Academy of Sciences, 60th October Anniversary Prospect 7a, Moscow 117312, Russia}
\author{Anatoly A. Semenov}
\affiliation{Department of Physics, Federal State Budget Educational Institution of Higher Education M.V. Lomonosov Moscow State University,
1(2), Leninskie Gory, GSP-1, 119991 Moscow, Russia}
\affiliation{Federal State Budget Educational Institution of Higher Education, M.V. Lomonosov Moscow State University, Skobeltsyn Institute of Nuclear Physics (SINP MSU), 1(2), Leninskie gory, GSP-1, 119991 Moscow, Russia}
\affiliation{Institute for Nuclear Research of the Russian Academy of Sciences, 60th October Anniversary Prospect 7a, Moscow 117312, Russia}
\date{\today}

\begin{abstract}
The detection of $\gamma$-ray candidates up to the energy of $\approx$13~TeV from the exceptionally bright $\gamma$-ray burst GRB 221009A by the Large High Altitude Air-shower Observatory (LHAASO) has raised considerable interest in the astrophysical community. The $\gamma$-ray dataset resulting from the LHAASO observations allows one to reconstruct the intrinsic spectrum of GRB 221009A with an unprecedented precision. This intrinsic spectrum reveals a downturn at the energy of several TeV (statistical significance $> 5 \sigma$), i.e. the reconstructed intensity is below the intensity expected for a power-law spectrum. We show that a significant TeV $\gamma$-ray component may be produced by neutrons from photohadronic interactions inside the fireball. These neutrons escape the fireball and interact with the surrounding matter, giving rise to a flux of electrons and positrons, eventually resulting in an observable flux of GeV--TeV synchrotron photons --- a high energy ``echo'' of the GRB prompt emission phase. Finally, we show that at multi-TeV energies the contribution of \mbox{$\gamma$ rays} from intergalactic electromagnetic cascades initiated by primary ultra high energy protons is severely limited during the early afterglow phase for the typical magnetic field strength in the intergalactic filaments above 1~nG.
\end{abstract}
\maketitle
\section{Introduction}

By the end of 2025, at least five $\gamma$-ray bursts (GRBs) have been detected in the very high energy (VHE, \mbox{$E > 100$~GeV}) domain, namely: 1) GRB 190114C \citep{MAGIC2019a,MAGIC2019b}, \mbox{2)~GRB~180720B} \citep{Abdalla2019b}, 3) GRB 190829A \citep{Abdalla2021}, \mbox{4)~GRB~201216C} \citep{Abe2023}, and 5) GRB 221009A \citep{LHAASO2023a,LHAASO2023b} (hereafter L23a,L23b). The observations of GRB 221009A are unique in more than one respect: this GRB was exceptionally bright \citep{Lesage2023,Williams2023,Burns2023} \footnote{GRB 221009A is sometimes referred to as ``the brightest of all time'' (BOAT) \citep{Burns2023}} and relatively nearby (redshift $z = 0.151$ \citep{deUgartePostigo2022,CastroTirado2022,Malesani2025}), while the principal observing detector in the VHE range (the LHAASO array \citep{Ma2022}) has an excellent sensitivity and a large field-of-view. The LHAASO array includes the LHAASO-WCDA and LHAASO-KM2A subdetectors sensitive to charged particles (thus not requiring clear moonless nights for observations in contrast to imaging atmospheric Cherenkov telescopes). More than $6.4 \times 10^{4}$ \mbox{$\gamma$ rays} from GRB 221009A were detected by the LHAASO observatory (L23a) during the first 3000~s after the Fermi-GBM trigger \citep{Lesage2023}; these events span a relatively wide energy range from 200~GeV to $\approx$13~TeV. For nine $\gamma$-ray candidates with the highest reconstructed energy, the chance probability of each event due to background is from 0.045 to 0.17 (L23b).

Primary $\gamma$ rays escaping the central GRB engine represent the so-called ``intrinsic'' spectrum; during the intergalactic propagation of the primary $\gamma$ rays this intrinsic spectrum is transformed due to the absorption on extragalactic background light (EBL) photons by means of the pair production (PP) process ($\gamma \gamma \rightarrow e^{+} e^{-}$) \citep{Nikishov1962,Gould1967}. Much effort was devoted to the study of various intergalactic $\gamma$-ray propagation models inspired by the detection of multi-TeV $\gamma$-ray candidates from GRB 221009A; these models sometimes include new physics effects such as photon-axion-like particle (ALP) mixing \citep{Troitsky2022,Galanti2023,Galanti2022,Carenza2022,Baktash2022,Wang2023,Gonzalez2023,Lin2023,Nakagawa2023,Zhang2022,AvilaRojas2024}, Lorentz invariance violation (LIV) \citep{Finke2023,Li2023,Li2024,Satunin2026}, or other exotic particles/effects beyond the Standard Model \citep{Smirnov2023,Balaji2023,Guo2023,Brdar2023,Huang2023}.

\begin{figure*}
\includegraphics[width=8.5cm]{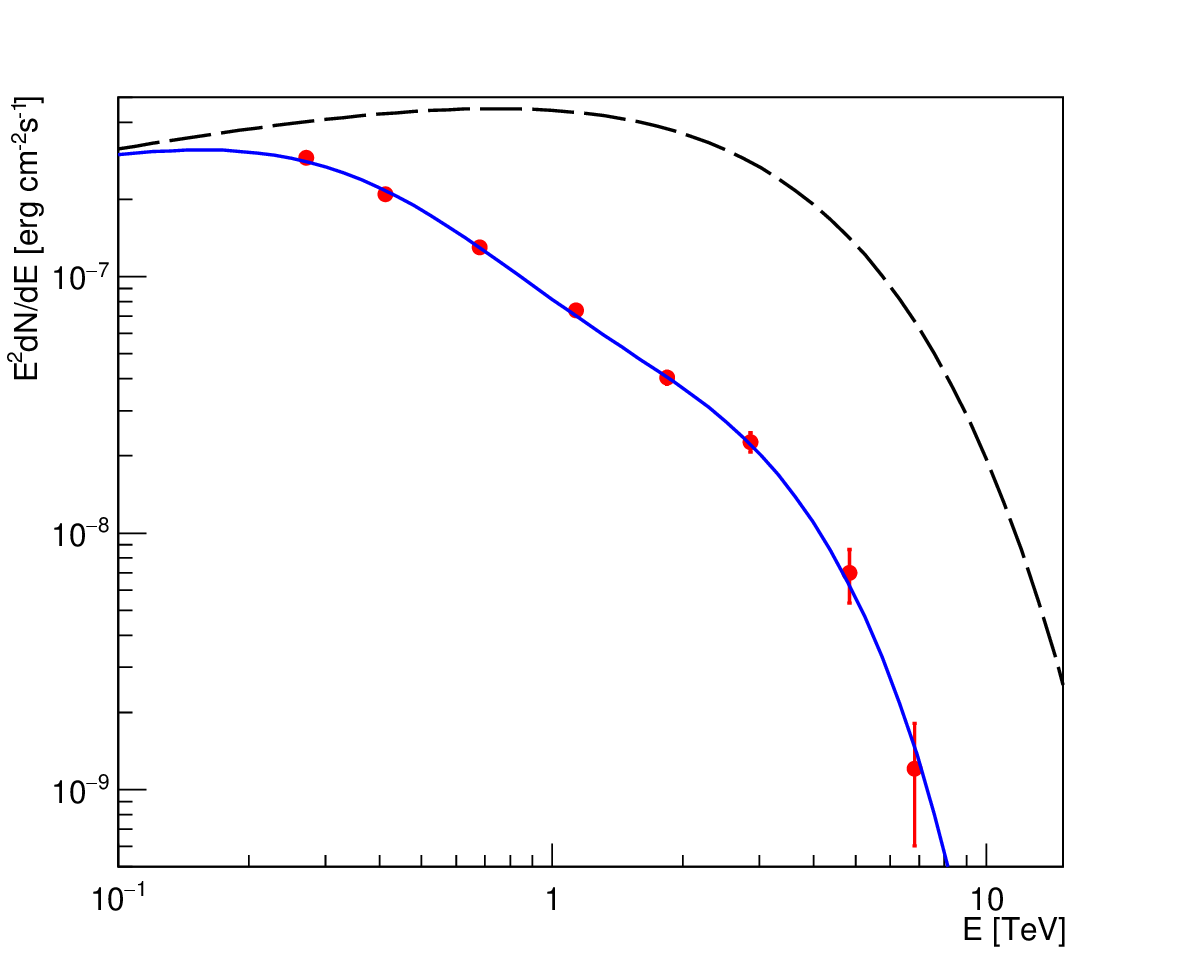}
\includegraphics[width=8.5cm]{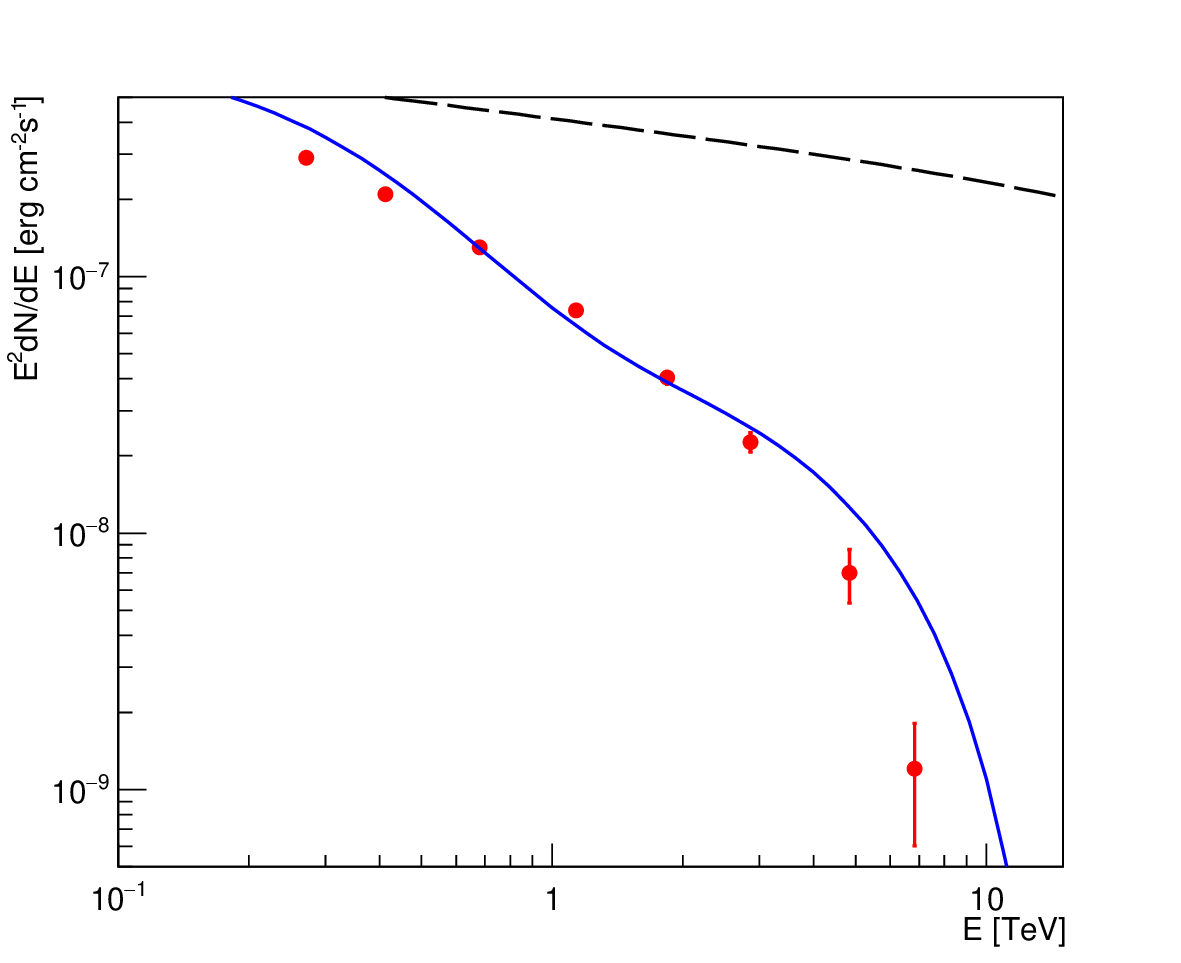}
\caption{The SED measured with the LHAASO-WCDA sub-detector for the time range of 326--900~s (red circles with statistical uncertainties), the observable model SED (blue curve), and the intrinsic SED (dashed black curve). The fit was obtained assuming the EPWL (left) and PWL (right) intrinsic spectral shapes. \label{fig:SED4}}
\end{figure*}

\begin{figure*}
\includegraphics[width=8.5cm]{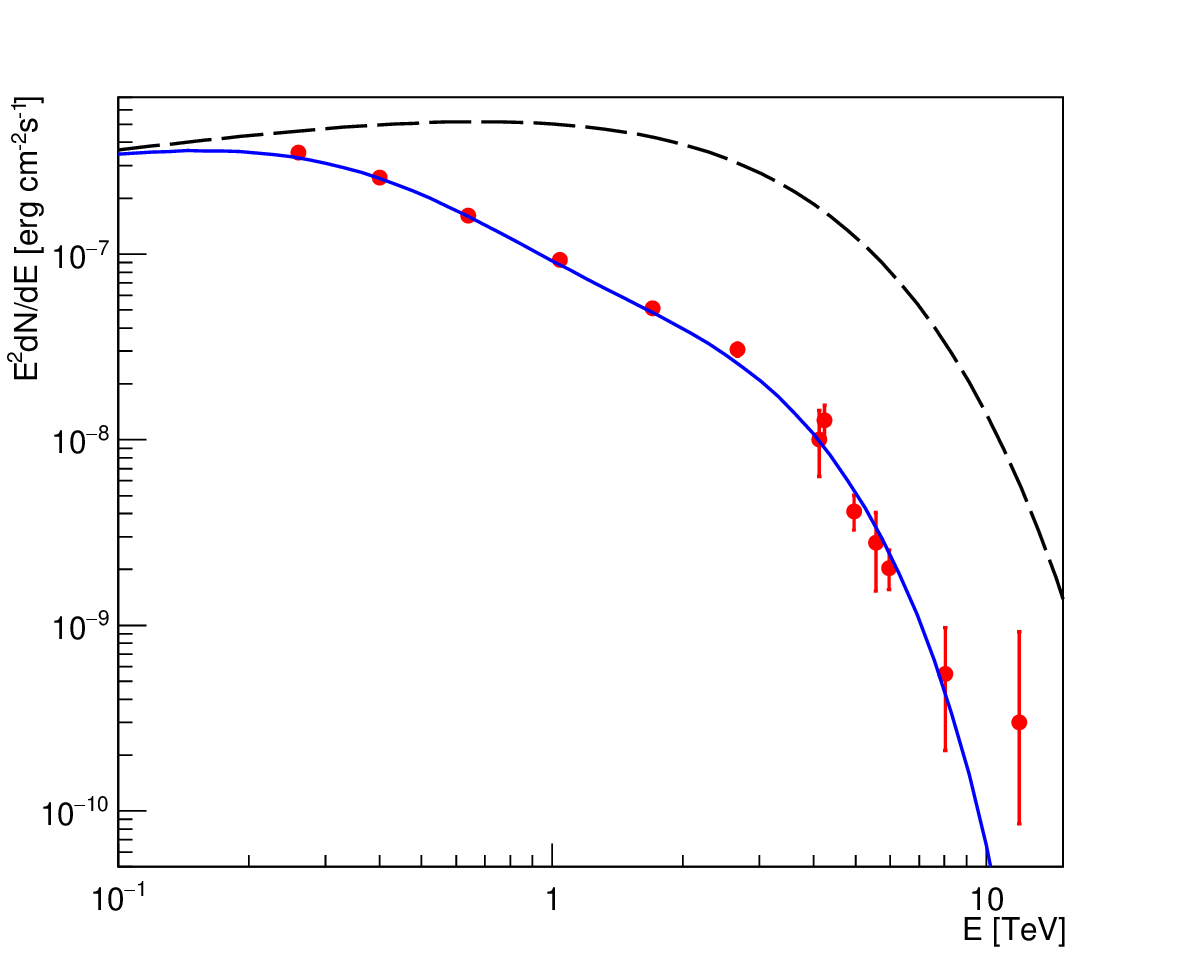}
\includegraphics[width=8.5cm]{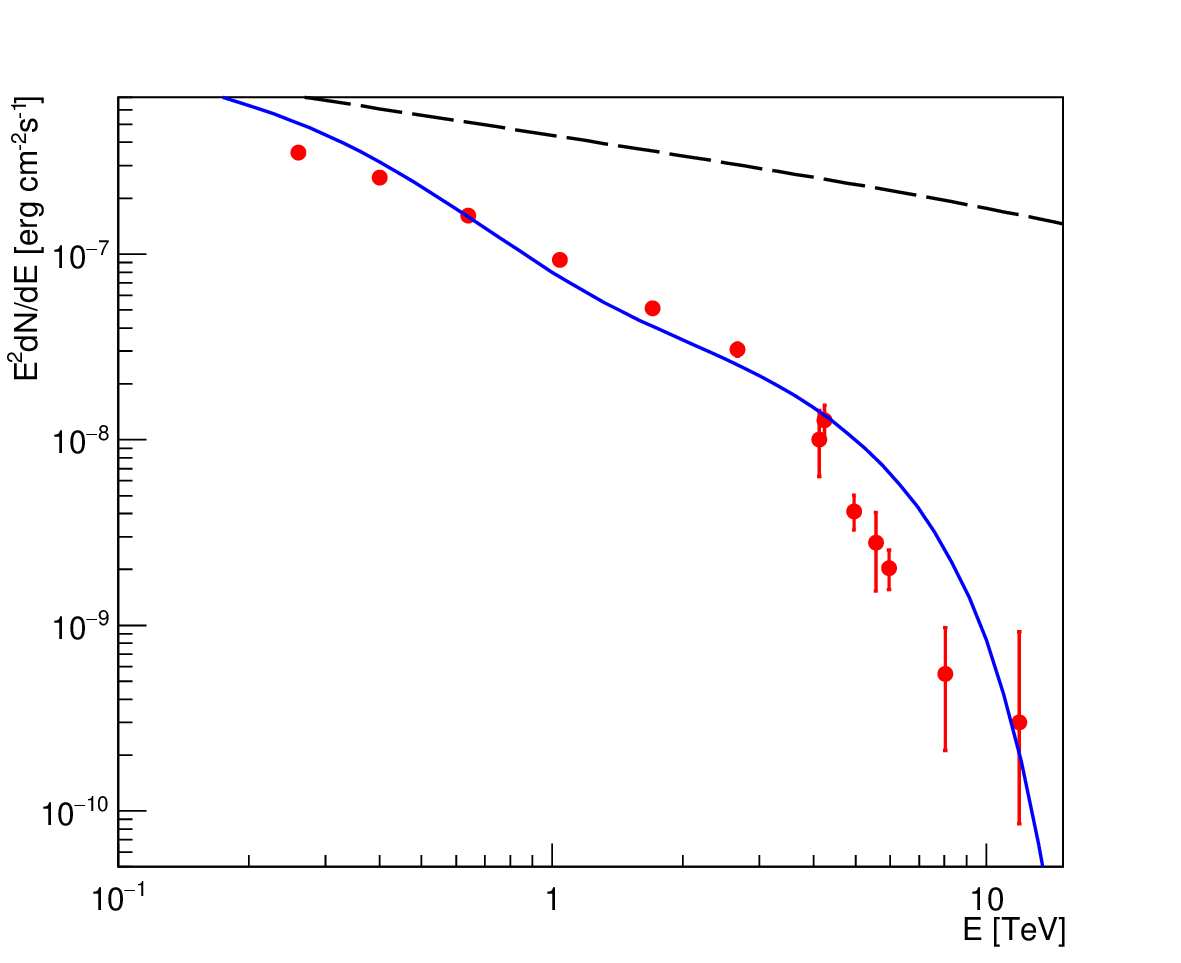}
\caption{Same as in Fig.~\ref{fig:SED4}, but for the combined LHAASO-WCDA and LHAASO-KM2A SED over the time range of 300--900~s. \label{fig:SED-Combined}}
\end{figure*}

In Sect.~\ref{sect:intrinsic} of this {\it Letter} we show that the reconstructed intrinsic $\gamma$-ray spectrum of GRB 221009A above the energy of several TeV is not compatible with a power-law shape and instead reveals a downturn. In Sect.~\ref{sect:neutrons} we discuss the characteristics of the GeV--TeV $\gamma$-ray echo created by the ultra high energy (UHE) neutrons escaping from the fireball of GRB 221009A. In Sect.~\ref{sect:ihcm} we consider the ``intergalactic hadronic cascade model'' (IHCM) which postulates a significant contribution from intergalactic electromagnetic (EM) cascades initiated by primary UHE protons or nuclei to the observable $\gamma$-ray spectrum. Some implication of our studies are discussed in Sect.~\ref{sect:discussion}. Finally, we conclude in Sect.~\ref{sect:conclusions}. Throughout this work we assume the following values of cosmological parameters: $H_{0} = 67.4$~km$\:$s$^{-1}$Mpc$^{-1}$, $\Omega_{m} = 0.315$ and $\Omega_{\Lambda} = 1 - \Omega_{m}$ \citep{Aghanim2020}.

\section{The shape of the intrinsic spectrum \label{sect:intrinsic}}

\begin{table}
\caption{Results for the G12 EBL model.} \label{tab:G12}
\begin{tabular}{| c | c | c | c c |}
\hline      
Time period [s] & p-value (EPWL) & p-value (PWL)          & Z-value (PWL) &\\
\hline                    
231--240        & 0.985          & 0.869                  & $<0$          &\\
240--248        & 0.163          & $8.70 \times 10^{-3}$  & $2.38 \sigma$ &\\
248--326        & 0.623          & $1.54 \times 10^{-16}$ & $8.17 \sigma$ &\\
326--900        & 0.804          & $1.28 \times 10^{-29}$ & $11.2 \sigma$ &\\
900--2000       & 0.746          & $1.21 \times 10^{-2}$  & $2.25 \sigma$ &\\
\hline                    
\end{tabular}
\end{table}

Let us assume that the intrinsic VHE $\gamma$-ray spectrum (the number of $\gamma$ rays $dN$ per unit energy $dE$, unit area $dA$, and unit time $dt$) of GRB 221009A follows a simple shape $dN/(dEdAdt) \propto E^{-\gamma}exp(-E/E_{c})$ (hereafter called the power-law with an exponential cutoff or EPWL for brevity), where $\gamma$ is the power-law index, and $E_{c}$ is the cutoff energy. We adopt the EBL model of Gilmore et al. \citep{Gilmore2012} (hereafter G12) and obtain a fit to the spectrum measured with LHAASO following the approach described in Sect. III of \citep{Dzhatdoev2020}. As an example, we take the spectrum measured with the LHAASO-WCDA sub-detector for the time range of 326--900~s (see Supplementary Information, Fig. S4C in L23a); hereafter all time ranges are measured since the Fermi-GBM trigger \citep{Lesage2023}. We note that the Fermi-GBM detector triggered on a precursor episode of GRB 221009A (for a discussion of precursors in GRBs see e.g. \citep{Lazzati2005}).

The measured spectral energy distribution (SED = $E^{2}dN/(dEdAdt)$; we will mark the SEDs presented in figures as $E^{2}dN/dE$ for simplicity) is shown in Fig.~\ref{fig:SED4} (left) together with the observable model SED and the intrinsic (reconstructed) SED; the latter reveals a cutoff at $E_{c}= 2.33$~TeV; the best-fit value of $\gamma$ is 1.68. The qua\-li\-ty of the fit is reasonably good (p-value $p = 0.804$; this quantity was calculated according to the prescriptions of \citep{Zyla2020}, Subsect. 40.3.2). In this particular case the p-value is greater than 0.5 ($p = 0.5$ corresponds to the statistical significance of $Z = 0$, i.e. a good fit without an overfitting). 

In Fig.~\ref{fig:SED4} (right) we present a similar fit, but for the case of a pure power-law (PWL for brevity) intrinsic spectrum; the best-fit value of $\gamma$ is 2.21. In this case the fit is very bad; $p = 1.28 \times 10^{-29}$ corresponding to $Z = 11.2 \sigma$. We conclude that the PWL intrinsic spectrum is not compatible with the LHAASO-WCDA dataset for GRB 221009A.

We present the p-values for the EPWL and PWL intrinsic spectral shapes for all five time ranges of the LHAASO-WCDA observations in Table~\ref{tab:G12}. In two cases the Z-value is in excess of 5~$\sigma$ indicating a strong incompatibility of the respective datasets with the hypothesis that the intrinsic spectra have a power-law shape. We do not insist on the exponential shape of the downturn in the intrinsic spectrum; for instance, this downturn may take a form of a curvature, a break, or a cutoff.

For the combined LHAASO-WCDA and LHAASO-KM2A spectrum (see Fig. 2B in L23b) and the time range of 300--900~s the results are presented in Fig.~\ref{fig:SED-Combined}. For the EPWL option $E_{c}= 2.06$~TeV, $\gamma = 1.67$, and $p = 0.294$, for the PWL option $\gamma = 2.35$ and $p = 1.22 \times 10^{-72}$ corresponding to $Z = 18 \sigma$. For the time range of 230--300~s the same values are $E_{c}= 2.91$~TeV, $\gamma = 1.96$, and $p = 0.672$ for the EPWL option and $\gamma = 2.43$, $p = 3.22 \times 10^{-23}$ corresponding to $Z = 9.9 \sigma$ for the PWL option.

p-values and Z-values for the EBL model of Saldana-Lopez et al. \citep{SaldanaLopez2021} (hereafter S21) are presented in Appendix~\ref{sect:tables}. We note that for the time range of 300--900~s and the combined LHAASO-WCDA and LHAASO-KM2A spectrum the statistical significance $Z = 9.0 \sigma$ is still higher than 5~$\sigma$, a commonly accepted threshold for a discovery in $\gamma$-ray astronomy.

\section{On the neutron beam model \label{sect:neutrons}}

In Fig. 7 of L23b, the LHAASO collaboration presents a comparison between the intrinsic spectrum of GRB~221009A reconstructed by them and a synchrotron self-Compton (SSC) model of this spectrum. In the time range of 300--900~s around the energy of several TeV the measured intensity is significantly above the model intensity. Thus, there is an indication for an additional component distinct from the SSC one contributing in this energy range to the observable spectrum.

The production of neutrons in photohadronic interactions inside the fireball of the GRB with the subsequent escape of these neutrons eventually results in a flux of GeV--TeV $\gamma$ rays mainly from synchrotron radiation of electrons and positrons produced in interactions of the escaping neutrons with the interstellar matter of the star-forming region (SFR) where the GRB is located \citep{Dzhatdoev2024}; Subsect. IV.B of \citep{Dzhappuev2025}. This ``neutron beam model'', usually invoked in active galactic nuclei (AGN) and GRB studies, has a long and interesting development history briefly covered in Appendix~\ref{sect:neutron-history}.

\begin{figure}
\vspace{0.1cm}
\includegraphics[width=8.5cm]{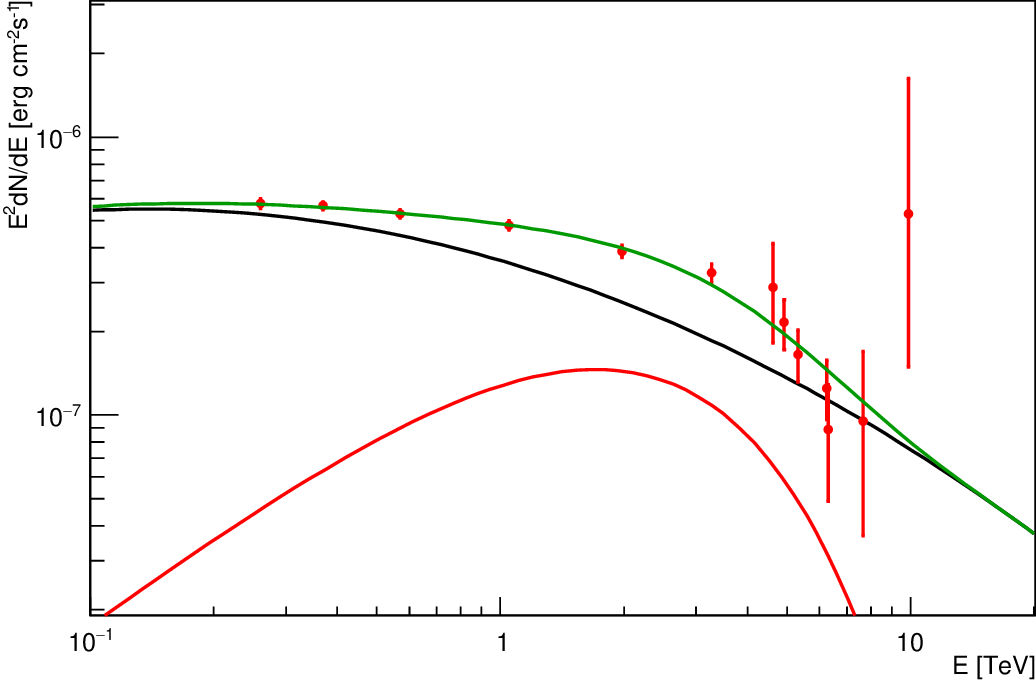}
\caption{The intrinsic SED according to Fig. 7B of L23b (red circles with statistical uncertainties), an approximation of their SSC model (black curve), an additional component (red curve), and the sum of the two latter components (green curve). \label{fig:Neutrons}}
\end{figure}

The intrinsic SED for the time range of 300--900~s calculated by the LHAASO collaboration is shown in Fig.~\ref{fig:Neutrons} together with their SSC model approximated with a log-parabolic function. To obtain a better fit to the LHAASO SED than the one provided with the SSC model, we include an additional spectral component $dN/(dEdAdt) \propto E^{-\gamma}exp(-E/E_{c})$ with $\gamma = 1.12$ and \mbox{$E_{c} = 1.53$~TeV}. The total model SED fits the data reasonably well. The total fluence of the additional (hard) component is:
\begin{equation}
S_{Hard} = \Delta t \int\limits_{E_{min}}^{E_{max}} E \frac{dN}{dEdAdt} dE = 2.17 \times 10^{-4} \; \frac{erg}{cm^{2}}, \label{eq1}
\end{equation}
where $\Delta t = 600$~s, $E_{min} = 100$~GeV, and \mbox{$E_{max} = 100$~TeV}.

The inelasticity coefficient for the photohadronic process at threshold is $f_{p \gamma} = 0.13$ \citep{Kelner2008}; neutrons are produced in $\approx 50$~\% of these interactions and carry the fraction of the primary proton energy $\approx 1 - f_{p \gamma} = 0.87$. Upper limits on the neutron fluence may be derived from the non-observation of muon neutrinos coincident in time and direction with GRB~221009A (see Appendix~\ref{sect:neutrino}). Assuming the same angular distribution for the neutrons and the accompanying prompt emission $\gamma$-rays of hadronic nature \footnote{this approximation is justified in the scenario with an isotropic target photon field internal to the fireball}, we estimate the maximum neutron fluence in the considered photohadronic scenario as follows \footnote{we note, however, that an additional neutron flux may come from the process of photodisintegration of nuclei, see Appendix~\ref{sect:neutron-history}}:
\begin{equation}
S_{n} \approx \frac{1-f_{p \gamma}}{f_{p \gamma}} S_{Tot} f_{h-max} = 7.7 \times 10^{-2} \; \frac{erg}{cm^{2}}, \label{eq2}
\end{equation}
where $f_{h-max} = 5.75 \times 10^{-2}$ is the maximum fraction of the prompt emission fluence due to photohadronic interactions (see Appendix~\ref{sect:neutrino}). Assuming that hadronic interactions of VHE neutrons and protons are identical due to the isotopic invariance of strong interactions, we estimate the fraction of the primary neutron energy channelled into secondary $\gamma$ rays $f_{\gamma} = 0.21$ \citep{Knapp1996} and the same quantity for electrons $f_{e} = 0.10$ \citep{Kelner2006}. The typical column density of the SFR is $N_{H} \ge 6 \times 10^{23}$~nucleon/cm$^{2}$ \citep{Krumholz2008}, the proton-proton inelastic cross section at 100~PeV is $\sigma_{pp} = 88$~mb \citep{Kelner2006}, corresponding to the proton-proton optical depth of $\tau_{pp} = \sigma_{pp} N_{H} \ge 5.3 \times 10^{-2}$, resulting in the synchrotron photon fluence of $S_{syn} = S_{n} \tau_{pp} f_{e} \ge 4.1 \times 10^{-4}$~erg/cm$^{2} \approx 1.9 S_{Hard}$. We conclude that the fluence of these synchrotron photons in the considered scenario is sufficient to fill the gap between the intrinsic SED and the SSC model in Fig.~\ref{fig:Neutrons}.

Following \citep{Dermer2004}, the temporal width of the emission pulse of the hard component may be estimated as \mbox{$\Delta t_{m} = R(1 - cos(\theta_{j}))/c$}, where $R$ is the distance from the central engine, and $\theta_{j} = 0.8^{\circ}$ is the jet half-opening angle (see L23a). The value of $R \approx 10^{4}$~AU = $1.5 \times 10^{17}$~cm \citep{Forrest1986} is defined by the radius of the cavity blown by the stellar wind in the SFR material. We obtain $\Delta t_{m} = 490$~s, not far from $\Delta t = 600$~s. We estimate the value of $E_{c}$ using a simple approximation for the sychrotron photon energy presented in \citep{Khangulyan2019}:
\begin{equation}
E_{cm} = \frac{6 \times 10^{7} \; eV}{1+z} \left( \frac{E_{e}}{1 \; PeV} \right)^{2} \left( \frac{B_{SFR}}{1 \; mG} \right), \label{eq3}
\end{equation}
where the electron energy is:
\begin{equation}
E_{e} \sim \frac{(1-f_{p \gamma})E_{p-max}^{\prime}D}{20} = 174 \; PeV \label{eq4}
\end{equation}
for protons accelerated in the rest frame of the fireball up to $E_{p-max}^{\prime} = 4$~PeV. We assume the magnetic field strength in the SFR $B_{SFR} \sim 1$~mG \citep{Crutcher1999} and obtain $E_{cm} = 1.58$~TeV, not far from the best-fit value $E_{c} = 1.53$~TeV. Finally, we note that the power-law index of the hard component $\gamma \approx 1.1$ is achievable in the framework of the considered scenario; for instance, the spectrum presented in Fig. 1 of \citep{Razzaque2009} is reasonably well approximated assuming $\gamma = 1.2$ and \mbox{$E_{c} = 8.8$~TeV}.

\section{The intergalactic hadronic cascade model \label{sect:ihcm}}

$\gamma$ rays from intergalactic EM cascades represent another kind of a high energy ``echo'' of the primary GRB emission. The case of a purely EM cascade initiated by primary $\gamma$ rays was considered in \citep{Dzhatdoev2023}; no signal from the cascade $\gamma$-ray echo was found and upper limits on its SED were presented in this paper. Here we consider the case of UHE proton-initiated EM cascades following \citep{Waxman1996}. For the proton energy $E_{p} > 10$~EeV the energy loss rate $-E_{p}^{-1}dE_{p}/dt$ at $z = 0$ exceeds $2 \times 10^{-10}$~yr$^{-1}$ (e.g. \citep{Berezinsky2006}, hereafter B06) corresponding to the energy loss length $d_{E-p} < 1.5$~Gpc. In this case the comoving distance to GRB~221009A $d_{c} = 647.1$~Mpc is comparable to $d_{E-p}$; therefore, the protons lose an appreciable fraction of their energy on their path from the source to the observer.

Secondary electrons, positrons, and $\gamma$ rays created in pair production acts (via the Bethe-Heitler process) and in photohadronic interaction initiate intergalactic EM cascades. A non-negligible fraction of these cascades is initiated relatively near to the observer (compared to the value of $d_{c}$) and, therefore, their observable spectra are relatively hard in the TeV energy domain (compared to the intrinsic spectrum absorbed on the EBL) \citep{Uryson1998}. We note that the IHCM for GRB~221009A was considered in \citep{Das2023,Mirabal2022}.

UHE protons initiate EM cascades on their way from the source to the observer, resulting in a spatially distributed EM cascade source function (injection term); the shapes of the observable spectra of these cascades become almost independent on the energy of the primary $\gamma$-ray or electron $E_{0}$ \citep{Berezinsky1975} when $E_{0} \ge 1$~PeV and the injection redshift $z_{g} \ge 5 \times 10^{-3}$, but still depend on the value of $z_{g}$ (``weak universality'', see \citep{Berezinsky2016,Dzhatdoev2017,Khalikov2021} for discussions and relevant numerical results). We present a simple semi-analytic approximation of the observable spectra of intergalactic EM cascades for various values of $z_{g}$ in Appendix~\ref{sect:cascade-appr}. 

A typical value of the proton deflection angle on a single intergalactic filament is $\delta = 3.1 \times 10^{-2} f_{BDE}$~rad, where
\begin{equation}
f_{BDE} = \left( \frac{B_{reg-fil}}{1 \; nG} \right) \left( \frac{d_{fil}}{1 \; Mpc} \right) \left( \frac{E_{p}}{30 \; EeV} \right), \label{eq5}
\end{equation}
$B_{reg-fil}$ is the strength of the regular magnetic field in the filament, and $d_{fil}$ is the diameter of the filament. Following \citep{Alcock1978}, we estimate the average time delay acquired by the protons due to their deflection inside the filaments on their way from the source to the observer:
\begin{equation}
\Delta t = \frac{d_{c}^{2}}{c \Delta d_{fil}} \frac{\delta^{2}}{12}, \label{eq6}
\end{equation}
where the average distance separating the neighboring filaments crossed by the proton $\Delta d_{fil} = d_{c}/N_{fil}$ and $N_{fil}$ is the average number of the filaments crossed by the proton on its way from the source to the observer.

The typical average strength of magnetic fields in extragalactic filaments at $z = 0$ is \mbox{$B_{fil} = 11-15$~nG} according to \citep{Carretti2025}; in addition, \citep{Vernstrom2021} find \mbox{$B_{reg-fil}= (0.05-0.15) B_{fil}$}. Assuming $B_{fil} = 10$~nG, $B_{reg-fil}= 0.1 B_{fil} = 1$~nG, $d_{fil} = 1$~Mpc (e.g. \citep{Yang2025}), and $E_{p} = 30$~EeV, we get $f_{BDE} = 1$. Furthermore, for most models considered in \citep{Tjemsland2024}, they estimate $\Delta d_{fil} = 15-25$~Mpc; assuming $\Delta d_{fil} = 20$~Mpc, we obtain the value of $\Delta t = 5.7 \times 10^{6}$~yr, which exceeds the LHAASO observation time range by eleven orders of magnitude. Therefore, most $\gamma$ rays produced by the primary protons are not observable with the existing $\gamma$-ray detectors.

\section{Discussion \label{sect:discussion}}

\subsection{Downturn in the intrinsic spectrum}

After the observation of GRB~221009A with LHAASO and the publication of an announcement to that effect \citep{Huang2022} (hereafter H22), including the detection of a $\gamma$-ray candidate with the reconstructed energy of 18 TeV, many studies have concluded that the conventional framework of ``the absorption and redshift-only model'', accounting for only the attenuation of the primary $\gamma$-ray flux on the EBL photons and adiabatic losses, is not sufficient to interpret the information communicated in H22 (e.g. \citep{Troitsky2022,Galanti2023,Baktash2022}).

Taking advantage of the datasets accompanying the refereed publications L23a and L23b, we demonstrate that the intrinsic spectrum of GRB~221009A does not have an excess at high energies w.r.t. a power law, but instead reveals a downturn. The fits to the spectra measured with LHAASO for the case of the EPWL intrinsic spectrum are good. Therefore, we conclude that it is not necessary to invoke any non-conventional intergalactic \mbox{$\gamma$-ray} propagation model in order to interpret the datasets presented in L23a and L23b. The LHAASO Collaboration have recently presented a new enhanced dataset of $\gamma$-ray-like events for GRB~221009A with new selection criteria allowing to lower the effective energy threshold \citep{Tang2025,Zhou2025}. After this dataset will hopefully be made publicly available, it would have become possible to reach a more robust conclusion about the shape of the intrinsic spectrum.

Assuming the SSC model, \citep{Derishev2019} consider the production of the observed sub-TeV $\gamma$ rays from GRB~190114C and find that inverse Compton scattering in this case typically occurs at the border between the Thomson and Klein-Nishina regimes. Therefore, one can expect a downturn in the spectrum of GRB~190114C at the energy of $\sim$1~TeV and possibly for GRB~221009A as well for comparable conditions. This is indeed supported by the SSC model presented in L23a and utilized in L23b. 

\subsection{The neutron beam model}

The neutron beam model provides an example of a multimessenger astrophysical scenario connecting the characteristics of the cosmic ray protons accelerated inside the fireball, the neutrinos produced in photohadronic interactions, and observable $\gamma$ rays. The component of $\gamma$ rays produced by the primary neutrons escaping the fireball (see Sect.~\ref{sect:neutrons}) is subdominant w.r.t. the SSC component, but might, nevertheless, have an appreciable impact on the shape of the high energy downturn in the spectrum.

For the case of $D = 10^{3}$ corresponding to the bulk Lorentz factor of the prompt phase $\Gamma = 500$ for the viewing angle $\theta_{v} = 0$ the non-observation of a neutrino signal from GRB~221009A with the IceCube Observatory severely limits the fraction of the prompt emission fluence due to photohadronic interactions: $f_{h} < 6$~\% (see Appendix~\ref{sect:neutrino}). The initial bulk Lorentz factor $\Gamma_{0} \approx \Gamma$ for GRB~221009A could be estimated following \citep{Liang2010} as $\Gamma_{0} = 182 \times (E_{iso}/10^{52} \; erg)^{p} = 1.1 \times 10^{3}$ for $p = 0.25$ and the isotropic-equivalent total emitted energy \mbox{$E_{iso} = 1.5 \times 10^{55}$~erg} \citep{An2023}. In this case (again for $\theta_{v} = 0$) $D = 2 \times \Gamma = 2.2 \times 10^{3}$, and the constraint on $f_{h}$ becomes significantly weaker: $f_{h} < 30$~\%. Furthermore, $p = 0.25 \pm 0.03$ \citep{Liang2010}; assuming $p = 0.28$, we get an even weaker constraint on $f_{h}$: $f_{h} < 46$~\%.

We note that some parameter values assumed in Sect.~\ref{sect:neutrons} are only order-of-magnitude estimates. For instance, the strength of the SFR magnetic field $B_{SFR}$ is poorly known. The value of $\Gamma$ is probably somewhere between 300 \citep{Gao2023} and 1400.

GRBs may accelerate protons up to the energy of $\sim$100~EeV \citep{Waxman1995,Vietri1995}; in particular, the cosmic ray acceleration ``in the fireball itself'' \citep{Milgrom1996} (i.e. during the prompt phase) is possible (see e.g. \citep{Pelletier2000,Vietri2003,Gialis2004,Globus2015} for discussions). The scenario presented in Sect.~\ref{sect:neutrons}, however, does not require the acceleration up to 100~PeV in the rest frame of the fireball (corresponding to $\sim$100~EeV in the host galaxy rest frame), but only up to $\sim$4~PeV in the fireball rest frame (corresponding to $\sim$4~EeV in the host galaxy rest frame).

\subsection{Time delay for $\gamma$ rays in EM cascades \\ in the framework of the IHCM}

Some EM cascades are initiated by the primary UHE protons before the protons get deflected in the first intergalactic filament they meet on their way from the source to the observer. In this case the time delay of the observable $\gamma$ rays may be significantly smaller than was estimated in Sect.~\ref{sect:ihcm} for those EM cascades that were initiated near the observer (i.e. when the primary proton has already typically undergone the deflection in many filaments).

The horizon energy of the observable $\gamma$ ray for a cascade initiated near the source at $z = 0.15$ is \mbox{$E_{h} \approx 550$~GeV} for the G12 EBL model, corresponding to the typical energy of the parent cascade electron $E_{ep} \sim 13$~TeV, and to the typical energy of the $\gamma$ ray producing that electron $E_{\gamma p} \sim 26$~TeV. Following \citep{Neronov2009} and assuming the strength of the extragalactic magnetic field (EGMF) in voids of the large scale structure $B_{void} = 10^{-18}$~G \footnote{this was the border value in \citep{Dzhatdoev2023}: the non-observation of the intergalactic EM cascade echo excludes the case of $B_{void} \le 10^{-18}$~G; the precise border value of $B_{void}$, however, may vary somewhat depending on the assumed EBL model, the assumptions about the shape of the intrinsic spectrum and other model assumptions}, we estimate the typical time delay for these observable (cascade) $\gamma$ rays at the $\gamma$-ray horizon energy as $\Delta t_{EM}(E_{h}) \sim 2 \times 10^{3}$~s. This is comparable to the width of the LHAASO observation time range.

In the framework of the IHCM, some broadening of the observable $\gamma$-ray pattern is expected, too. This effect was discussed in \citep{Khalikov2021}.

\section{Conclusions \label{sect:conclusions}}

In the present {\it Letter}, we have considered the shape of the very high energy $\gamma$-ray spectrum of GRB 221009A, including the multi-TeV $\gamma$-ray candidates registered from this source. The power-law shape of the intrinsic spectrum is not compatible with the LHAASO-WCDA and the combined LHAASO-WCDA and LHAASO-KM2A datasets for GRB 221009A; the corresponding statistical significance exceeds $5 \sigma$. The very high energy spectrum of GRB 221009A is well described by a sum of two components: 1) the ``standard'' SSC component, well-fitted by a log-parabolic spectrum and 2) the ``anomalous'' component produced by the ultra high energy neutrons escaping the fireball freely, well-described by a hard ($\gamma \approx 1$) spectrum with a cutoff at $\approx$1.5~TeV. This hard, ``anomalous'' component is essentially a very high energy $\gamma$-ray echo of the GRB prompt phase. In the framework of the intergalactic hadronic cascade model, the typical delay of the observable cascade $\gamma$ rays from electromagnetic cascades initiated near the observer is by many orders of magnitude greater than the LHAASO observation time range. Therefore, most $\gamma$ rays produced in this scenario are not observable with the existing $\gamma$-ray telescopes.

\begin{acknowledgments}
We acknowledge helpful discussions with E.I.~Podlesnyi. All graphs in this work were produced with the ROOT software toolkit \cite{Brun1997}. This work was supported by the Russian Science Foundation, grant no. \mbox{22-12-00253-P}.
\end{acknowledgments}
\bibliography{GRB221009A-MultiTeV}
\appendix

\section{Results for the S21 EBL model \label{sect:tables}}

We present the p-values for the EPWL and PWL intrinsic spectral shapes for all five time ranges of the LHAASO-WCDA observations and the S21 EBL model in Table~\ref{tab:S21}. For the combined LHAASO-WCDA and LHAASO-KM2A spectrum and the EPWL intrinsic spectral shape, the p-value is greater than 0.18 for both time windows. For the PWL intrinsic spectral shape, the p-values for the time ranges of 230--300~s and 300--900~s are $3.71 \times 10^{-5}$ ($Z= 4.0 \sigma$) and $8.86 \times 10^{-20}$ ($Z= 9.0 \sigma$), respectively.

\begin{table}
\caption{Results for the S21 EBL model.} \label{tab:S21}
\begin{tabular}{| c | c | c | c c |}
\hline      
Time period [s] & p-value (EPWL) & p-value (PWL)          & Z-value (PWL) &\\
\hline                    
231--240        & 0.931          & 0.919                  & $<0$          &\\
240--248        & 0.096          & $2.30 \times 10^{-2}$  & $2.00 \sigma$ &\\
248--326        & 0.309          & $2.21 \times 10^{-6}$  & $4.59 \sigma$ &\\
326--900        & 0.792          & $2.43 \times 10^{-8}$  & $5.46 \sigma$ &\\
900--2000       & 0.865          & 0.366                  & $0.34 \sigma$ &\\
\hline                    
\end{tabular}
\end{table}

\section{The historical development \\ of the neutron beam model \label{sect:neutron-history}}

In 1977, V.S. Berezinsky and O.F. Prilutskii proposed a mechanism for escape of cosmic ray protons from Galactic sources via the neutron production in hadronuclear interactions \citep{Berezinskii1977} (hereafter B77). In 1978, D. Eichler and P.J. Wiita considered photohadronic interactions of protons accelerated in AGN, resulting in the formation of neutron beams \citep{Eichler1978} (hereafter E78). We note that the scenario of B77 did not become a very frequently invoked one; however, in combination with the photohadronic mechanism of neutron production proposed in E78, this scenario might be helpful in interpreting the recently published $\gamma$-ray dataset on the microquasar Cygnus X-3 \citep{LHAASO2025} together with the cosmic ray datasets on the proton (or proton + Helium) spectrum in the energy range of 10--100~PeV (e.g. \citep{Apel2013a,Aartsen2019,Apel2013b,Antonov2015}). Furthermore, the neutron beam model was proposed for the case of the microquasar SS~433 \citep{Escobar2025,Fargion2025}.

The same photohadronic mechanism of neutron beam formation (as in E78), again for the case of AGN, was further elaborated on in \citep{Kirk1989}; in addition to the $\gamma$-ray production, the authors consider the X-ray production via synchrotron radiation of electrons and positrons from pair production acts. The process of photodisintegration of nuclei is considered in \citep{Tkaczyk1994}; in this case the threshold for the neutron production mechanism is lower than for the scenario of E78.

Photohadronic interactions of PeV--EeV protons accelerated in blazar blobs may lead to the formation of neutron beams (and eventually to the production of VHE $\gamma$ rays) \citep{Atoyan2003}. The same mechanism was applied to the case of the MeV-GeV $\gamma$-ray emission from GRB 941017 \citep{Dermer2004}. The energy of synchrotron $\gamma$ rays radiated by electrons and positrons from hadronuclear interactions of neutrons escaping the GRB's fireball may be as high as 1~TeV or even higher; for instance, \citep{Razzaque2009} find that the peak of the SED of this component is located at the energy of $\approx$10~TeV. We note that the upper limit on the energy of synchrotron photons radiated by shock-accelerated electrons (e.g. \citep{Kumar2012}) does not apply for secondary charged particles, for instance, for electrons produced in hadronuclear interactions.

\section{Neutrino constraints on the neutron beam model \label{sect:neutrino}}

\begin{figure}
\vspace{0.1cm}
\includegraphics[width=8.5cm]{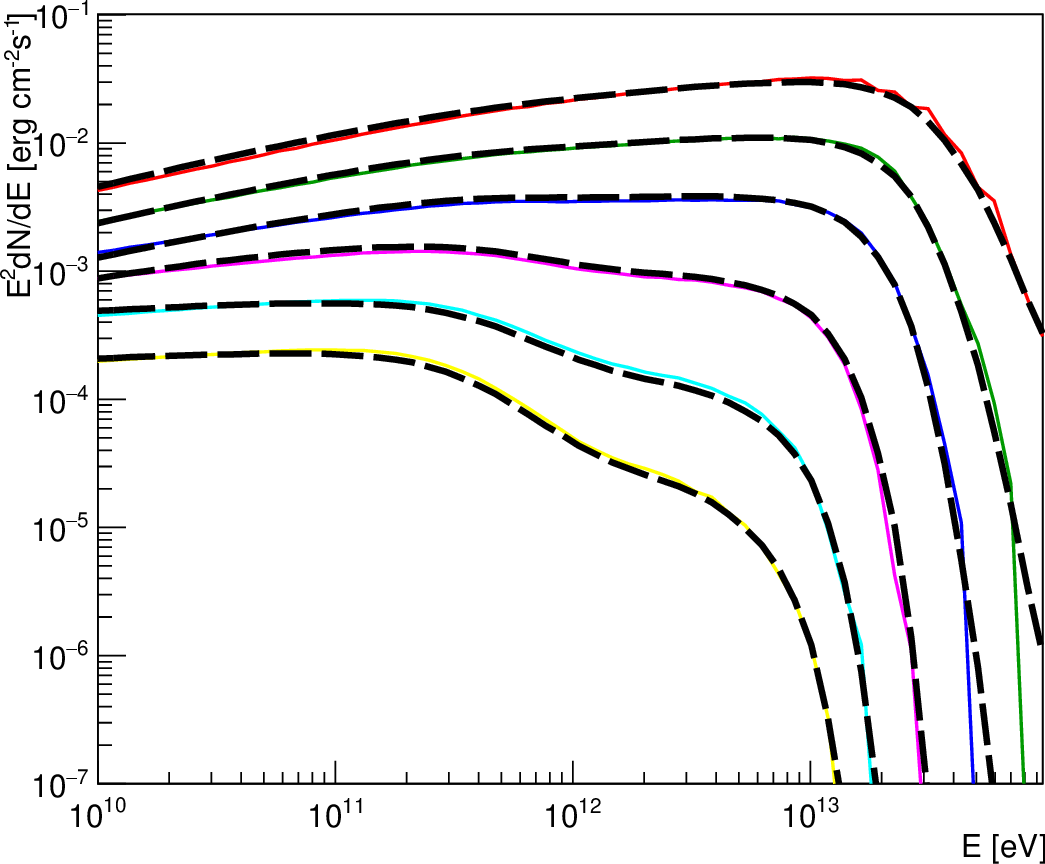}
\caption{Average observable SEDs of intergalactic EM cascades from 1~PeV $\gamma$ rays (solid curves) together with their approximations (dashed thick black curves). \label{fig:Appr}}
\end{figure}

The characteristic proton energy threshold for the photopion interaction in the rest frame of the fireball is: 
\begin{equation}
E_{p-thr}^{\prime} = \frac{m_{p} E_{\Delta} D}{E_{Peak}(1+z)} = 85.6 \; TeV, \label{eqc1}
\end{equation}
where $m_{p} = 938.272$~MeV is the proton mass, $E_{\Delta} = 315$~MeV is the energy of the photon in the rest frame of the proton corresponding to the maximum in the $p \gamma$ cross section (e.g. \citep{Kossov2002}) \footnote{the value of $E_{\Delta}$ may vary from 315 MeV to 340 MeV depending on the assumed approximation of the $p \gamma$ cross section}, the Doppler factor of the fireball $D = 10^{3}$ and the peak energy in the observed prompt emission SED $E_{Peak} = 3$~MeV \citep{Frederiks2023}. This corresponds to the characteristic energy of the observable neutrinos
\begin{equation}
E_{\nu} = \frac{E_{p-thr}^{\prime}D}{20(1+z)} = 3.72 \; PeV. \label{eqc2}
\end{equation}
The total fluence of the prompt emission from GRB~221009A is $S_{Tot} \approx 0.2$~erg/cm$^{2}$ = \mbox{1.25$\times 10^{11}$~eV/cm$^{2}$} \citep{Burns2023}. Assuming that the neutrinos and $\gamma$ rays have approximately equal luminosity (this approximation is valid if both the muons and pions produced in photohadronic interactions decay before radiating an appreciable part of their energy through synchrotron radiation), we estimate the average expected number of the observable muon neutrinos in the IceCube array as follows:
\begin{equation}
N_{\nu-avg} = \frac{f_{h}S_{Tot}A}{3 E_{\nu}} = 3.0 \frac{f_{h}}{5.75 \times 10^{-2}}, \label{eqc3}
\end{equation}
where $f_{h}$ is the fraction of the prompt emission fluence due to photohadronic interactions \footnote{with the subsequent development of electromagnetic cascades} and the IceCube effective area for $E_{\nu} = 3.7$~PeV is $A = 4.66 \times 10^{2}$ m$^{2}$ \citep{Abbasi2023}. The non-observation of a significant neutrino flux implies $N_{\nu-avg} < 3$ \footnote{on the other hand, $N_{\nu-avg} = 1$ does not necessarily imply the non-observation of a neutrino signal due to statistical fluctuations of the number of detected neutrinos}; therefore, we estimate the maximum value of $f_{h-max} = 5.75 \times 10^{-2}$.

\section{Approximation for the intergalactic EM cascade spectrum in the universal regime \label{sect:cascade-appr}}

Average observable SEDs of intergalactic EM cascades initiated by 1~PeV $\gamma$ rays for several narrow ranges of the injection redshift $z_{g}$ are shown in Fig.~\ref{fig:Appr}. These SEDs were calculated with the \mbox{ELMAG} code \citep{Kachelriess2012} version 3.01 \citep{Blytt2020} for the case of the G12 EBL model. In total, $10^{5}$ SEDs for the range of $0 < z_{g} < 0.3$ with a uniform distribution on $z_{g}$ were obtained. We select the following ranges of $z_{g}$ that correspond to solid curves in Fig.~\ref{fig:Appr} : $4 \times 10^{-3}$ -- $6 \times 10^{-3}$ (average $z_{g-av} = 5 \times 10^{-3}$, red curve),  0.008--0.012 ($z_{g-av} = 0.01$, green curve), 0.015--0.025 ($z_{g-av} = 0.02$, blue curve), 0.045--0.055 ($z_{g-av} = 0.05$, magenta curve), 0.095--0.105 ($z_{g-av} = 0.1$, cyan curve), 0.145--0.155 ($z_{g-av} = 0.15$, yellow curve). The array of the SED values $F_{i}$ for the top (red) curve was normalized to satisfy the relation $\sum_{i=0}^{N}{F_{i}} = 1\;erg\:cm^{-2}s^{-1}$; other arrays were first normalized in the same way and then all the values in them were multiplied by the factor of $10^{-0.4\cdot j}$, where $j$ is the number of the array.

Assuming weak universality and following \citep{Dzhatdoev2021}\footnote{in \citep{Dzhatdoev2021}, we performed a study of internal EM cascade development in extreme TeV blazars using an approximation similar to eq.~\ref{eq-casc}}, we approximate the average observable spectrum of intergalactic EM cascades as a smoothly broken power law with a high energy cutoff:
\begin{equation}
\frac{dN}{dE} \propto E^{-\gamma_{1}} \left[ 1 + \left( \frac{E}{E_{x}(z_{g})} \right)^{\varepsilon} \right]^{-\frac{\gamma_{2}-\gamma_{1}}{\varepsilon}}e^{-\tau(E,z_{g})}, \label{eq-casc}
\end{equation}
where $\gamma_{1} \approx 1.5-1.6$ \citep{Berezinsky2016,Dzhatdoev2017}, $\gamma_{2} \approx 1.9-2.0$ \citep{Berezinsky2016}, $\varepsilon$ defines the sharpness of the break, $\tau(E,z_{g})$ is the optical depth defined by the assumed EBL model. The transition between the segments with the power-law indices $\gamma_{1}$ and $\gamma_{2}$ occurs at $E_{x}(z_{g}) = (4/3)\gamma_{e}^{2}E_{cmb}$, where the cascade electron Lorentz factor $\gamma_{e} = E_{e-X}/m_{e}$, $E_{e-X} \approx E_{h}/2$, $E_{h}$ is the $\gamma$-ray horizon energy defined by the condition $\tau(E_{h},z_{g})=1$ and $E_{cmb} = 6.4 \times 10^{-4}$~eV is the characteristic energy of a CMB photon. Dashed thick black curves in Fig.~\ref{fig:Appr} are shown for the best-fit values of $\gamma_{1} = 1.53$, $\gamma_{2} = 1.97$, and $\varepsilon = 0.56$.
\end{document}